\documentclass[12pt,preprint]{aastex}
\usepackage{emulateapj5,natbib,graphics} 

\newcommand{\eqb}{\begin{eqnarray}}
\newcommand{\eqe}{\end{eqnarray}}
\newcommand{\eqen}{\nonumber\end{eqnarray}}

\def\chandra{{\it Chandra}}

\def\b2{B2 0902+34}

\def\kev{{\rm\thinspace keV}}

\def\ks{{\rm\thinspace ks}}

\def\c{\hbox{$\rm\thinspace cts$}}

\def\ergps{\hbox{$\rm\thinspace erg~s^{-1}$}}

\def\kmpspmpc{\hbox{$\rm\thinspace km~s^{-1}~Mpc^{-1}$}}

\def\erg{{\rm\thinspace erg}}

\def\keV{{\rm\thinspace keV}}

\def\s{{\rm\thinspace s}}

\def\ergps{\hbox{$\erg\s^{-1}\,$}}

\newcommand{\gtsim}{\mbox
{{\raisebox{-0.4ex}{$\stackrel{>}{{\scriptstyle\sim}}$}}}}
\newcommand{\ltsim}{\mbox
{{\raisebox{-0.4ex}{$\stackrel{<}{{\scriptstyle\sim}}$}}}}

\slugcomment{to appear in ApJ Letters}

\shorttitle{Low energy cutoff}
\shortauthors{Blundell et al}

\begin{document}

\title{\bf Discovery of the low-energy cutoff in a powerful \\ giant radio galaxy}
\author{Katherine M.\  Blundell\altaffilmark{1},
  A.C. Fabian\altaffilmark{2}, Carolin S. Crawford\altaffilmark{2},
  M.C. Erlund\altaffilmark{2},\newline  \& Annalisa Celotti\altaffilmark{3}\\    
} 

\altaffiltext{1}{University of Oxford, Astrophysics, Keble
  Road, Oxford, OX1 3RH, U.K.}
\altaffiltext{2}{Institute of Astronomy, Madingley Road,
    Cambridge, CB3 0HE, UK}
\altaffiltext{3}{International School for Advanced
    Studies, SISSA/ISAS, via Beirut 2-4, I-34014 Trieste, Italy}

\begin{abstract}
  The lobes of radio galaxies and quasars, fed by jets and hotspots,
  represent a significant, and currently ill-constrained, source of
  energy input into the inter-galactic medium (IGM).  How much energy
  is input into the IGM depends on the {\em minimum} energy to which
  the power-law distribution of relativistic particles is accelerated
  in the hotspots.  This has hitherto been unknown to within three
  orders of magnitude.  We present direct evidence for the discovery
  of this low-energy cutoff in the lobe of a Mpc-sized radio galaxy
  via the existence of extended X-ray emission, inverse-Compton
  scattered from {\em aged} radio plasma, and its separation by
  80\,kpc from regions containing {\em freshly accelerated} plasma
  from the hotspot.  The low-energy cutoff of $\gamma \sim 10^4$ in
  the hotspot is higher than previously thought, but reconciles
  discrepancies with magnetic field estimates which had been
  systematically lower than equipartition values.  The inverse Compton
  scattering of the spent synchrotron plasma is at the expense of
  cosmic microwave background (CMB) photons; we comment on the
  importance of such giant radio galaxies as contaminants of CMB
  anisotropies.
\end{abstract}

\section{Introduction}
\label{sec:intro}
Knowledge of the energy input into the IGM from a powerful radio
galaxy is necessary to understand the feedback of radio jets and lobes
on their environments and hence their influence on cosmic structure
formation. Estimates of this energy have been used to attempt
\citep{War98,Cel98,Cel03} to discriminate between whether the
constituents of jet plasma are electron-positron (light jets) or
electron-proton (heavy jets), which has considerable implications for
the jet-production mechanism at the supermassive black holes from
which the jets emanate.  To calculate the energy stored in the radio
lobes, it is necessary to integrate over the entire energy
distribution of the particles in the plasma which constitutes the
lobes.  The spectral shape, constrained by observations, is often a
good approximation to a power-law meaning that there are very few high
Lorentz factor ($\gamma$) particles and many orders of magnitude more
low-$\gamma$ particles.  The lower limit of this energy distribution,
hereafter $\gamma_{\rm min}$, was hitherto barely constrained by
observations or theory \citep{Car91,Har01} and is thus crucial for even
an approximate estimate of the energies involved.  A considerable
range of values for $\gamma_{\rm min}$ is covered by different authors
who are forced to make an arbitrary assumption (e.g.\ $\gamma_{\rm
min} = 1$ is used by \citet{Har05} and also by \citet{Kai97},
$\gamma_{\rm min} = 10$ is used by \citet{Cro05}, $\gamma_{\rm min} =
100$ is used by \citet{Car91}, 1000 is taken by \citet{War98}).  This
important point is actually rather obscured by the fact that many
authors state an assumed minimum frequency $\nu$ rather than a minimum
$\gamma$, following \citet{Mil80}.

\section{Comparison of X-ray and radio emission}
\label{sec:obs}
The \chandra\ X-ray satellite is revolutionizing studies of radio
galaxies and quasars since it is now possible to resolve extended
X-ray emission from hotspots, jets and lobes.  An extremely useful
web-site documents discoveries to
date.\footnote{http://hea-www.harvard.edu/XJET/} While studies of jets
and hotspots in these objects reveal much about speeds and magnetic
fields, our focus in this paper is on the radiation from the lobe
regions (i.e.\ plasma that has been output from the hotspot).  We
proposed for \chandra\ observations of some giant radio galaxies as
part of a study to investigate the population of relativistic
particles in their lobes having {\em lower} Lorentz factors than can
be probed by radio observations of synchrotron emission.  In all
standard models of radio source evolution \citep{Bal82,Kai97,Blu99}
longer (i.e.\ more extended) sources are older than shorter sources;
as such, giant radio galaxies should be expected to have older plasma
(having lower Lorentz factors than freshly accelerated plasma which
has suffered less losses) than shorter sources.

A particularly important target amongst these is the powerful
classical double \citep[FRII][]{Fan74} radio galaxy 6C\,0905+3955
(also known as 4C39.24), at the relatively high-redshift of $z = 1.88$
\citep{Law95}. This object has a physical size projected on the plane
of the sky of 945\,kpc [using the assumed cosmology of $H_{0} =
71$\kmpspmpc\, $\Omega_{\rm M}=0.27$ and $\Omega_{\Lambda} = 0.73$ and
its measured angular size of 111$^{\prime\prime}$].

\begin{figure*}
  \centering
\vbox{
  \includegraphics[width=0.95\columnwidth]{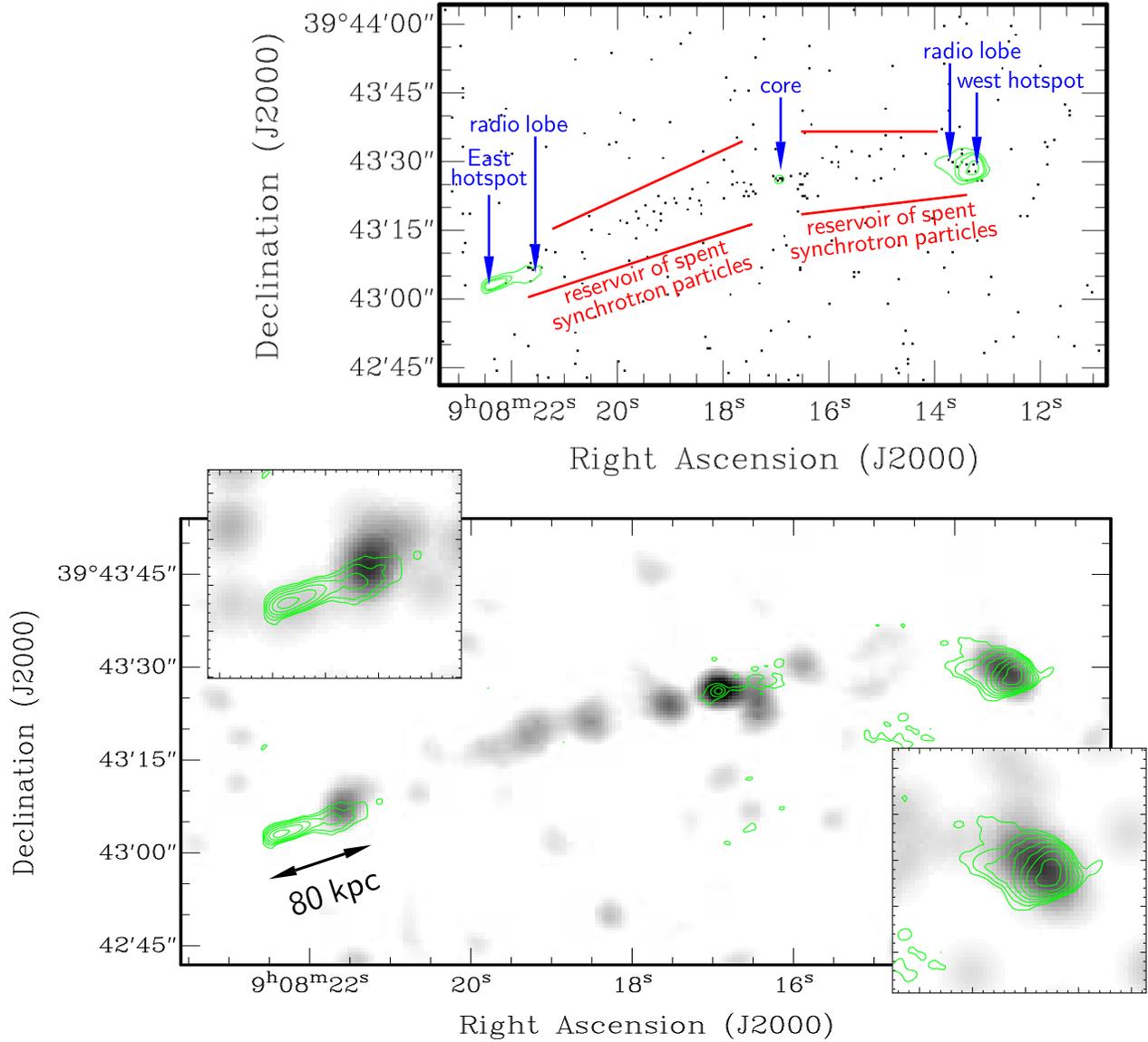}
  }
\caption{ \label{fig:overlay} Upper panel: greyscale is unsmoothed
  X-ray (0.3-3\keV) emission from the giant radio galaxy
  6C\,0905+3955; green contours are of radio emission at 1.4\,GHz from
  archival observations with the A-, B- and C-configurations of the
  VLA (the lowest contour is 0.6\,mJy/beam).  Blue labels indicate the
  nomenclature used for the radio emitting parts of the source and
  pairs of red lines indicate regions containing spent synchrotron
  lobe material having Lorentz factor particles that are too low to
  radiate at radio wavelengths but sufficient to inverse Compton
  scatter CMB photons to X-ray energies.  Lower panel: greyscale, as
  upper panel, but smoothed by a four-pixel Gaussian; contours of the
  same radio emission at 1.4\,GHz (the lowest contour is
  0.4\,mJy/beam).  }
\end{figure*}

The radio emission from 6C\,0905+3955, shown in
Figure\,\ref{fig:overlay}, and indeed at 408\,MHz in figure\,1a of
\citet{Law95} which closely resembles our image at 1.4\,GHz,
represents the entireity of the radio emission at these frequencies:
the radio lobes seen just inward of the compact hotspots, together
with the core, have the same silhouette at low frequency \citep{Law95}
as our high frequency image in Figure\,\ref{fig:overlay} shows.
The 408-MHz MERLIN image of \cite{Law95} demonstrably does not
undersample much smooth extended emission because the total flux
density measured from the MERLIN 408-MHz image is the same as the
measured total flux density of 940\,mJy measured by the Bologna
408-MHz survey \citep{Fic85} which measures integrated flux density.
Moreover, comparison of the integrated flux density from our 1.4-GHz
image ($266 \pm 3$ mJy) and that from the NVSS observations ($260 \pm
4$ mJy) by \citet{Con98} is consistent with no radio flux density
having been lost because of interferometric undersampling in the
extended VLA configurations.

The \chandra\ X-ray satellite observed 6C\,0905+3955 on 2005 March 02:
the X-ray emission from this object is well-detected in 19.68\ks\ with
about 114\c\ in total (all counts quoted are background subtracted),
and is spread in a linear fashion contained within the giant radio
structure (Figure\,\ref{fig:overlay}).  The (background subtracted)
counts are divided between: the lobe which contains 81\c\ and the
nucleus which contains 11\c, a region with 15\c\ associated with the
western radio hotpot and a region containing 9\c\ which is towards the
eastern radio hotspot, but actually 80\,kpc from it in the direction
of the nucleus.  There is no indication of any host cluster or group
in the relatively short exposure.  

Inspection of images in different energy bands indicates that the
nucleus or core is hard and the remainder (i.e.\ the emission from
between the hotspots but excluding the nucleus) is soft.  Despite the
low count rates, crude spectra are extracted and summarised in
Table\,\ref{tab:results}.
 
The unsmoothed image (upper panel, Fig\,\ref{fig:overlay}) shows that
the X-ray emission just to the east of the core is resolved transversely
with a width of $\sim 40$\,kpc, larger than the transverse width of
the radio lobe seen rather further to the East, and is thus most
un-jet-like.  The X-ray emission does not in any way resemble the fine
and slender collimated jet emission which has been observed in FR\,I
radio sources and low-redshift FR\,II
sources\footnote{http://hea-www.harvard.edu/XJET/} and so strongly
suggests that this X-ray emission arises from the same plasma that at
earlier times would have been radio-emitting lobe material.

Comparison of the X-ray and radio structures seen in
Figure\,\ref{fig:overlay} shows that the vast majority of the X-ray
emission seems to be associated with the regions {\em inside} the
radio lobes rather than spatially coincident with them (leaving aside
the nucleus where different physics is at play).  These regions are
where the hotspots would have {\em previously} passed through,
depositing then freshly accelerated plasma, on their journey outward
from the central nucleus of the active galaxy.  This plasma would have
radiated synchrotron emission for at most a few $10^7$ years at
observable radio wavelengths \citep{Blu00}, and is subject to energy
losses from four principal means \citep{Blu99}: i) adiabatic expansion
losses as the plasma escapes from the high pressure hotspot into the
lower pressure lobe, ii) adiabatic expansion losses as the
over-pressured lobe itself expands into the IGM, iii) synchrotron
losses and iv) inverse Compton losses off the CMB.  Simple
calculations show that (i) means that the Lorentz factors of all
particles will reduce by a factor of typically ten \citep{Sch68,Blu99}
after exiting the compact hotspot and expanding into the lobe and (ii)
means there will be a continuous reduction in their Lorentz factors by
an amount which depends on the subsequent fractional increase in
volume of the lobe.  [Note that (iii) and (iv) make only a slight
difference to the low-$\gamma$ particles since these losses depend on
$\gamma^2$.]  Thus, whatever is the minimum Lorentz factor
($\gamma_{\rm min}$) resulting from acceleration in the hotspot, it
will give rise to a $\gamma_{\rm min}$ which is at least an order of
magnitude lower in the lobes --- we now establish what these values of
$\gamma_{\rm min}$ are in these two locations.

\section{Relevant Lorentz factors}
\label{sec:gamma}

To suggest that the X-ray emission seen to the east of the nucleus is
synchrotron emission, in a region where there is negligible radio
emission even at 408\,MHz (see previous section) would require an
unprecedented spectrum of relativistic particles having significant
number density with Lorentz factors of $10^7$ and rather less with
$10^4$, which would in require an unprecedented acceleration mechanism;
therefore, it is assumed that there is no particle acceleration
taking place to cause this.  A much more conservative explanation is
the process of inverse-Compton scattering of CMB photons (ICCMB) by
relativistic electrons widely believed to be in the lobes of radio
galaxies, which we now explore.  The usual assumption is made
throughout that the plasma which comprises the lobe (or what used to
radiate at radio wavelengths as lobe material) is not moving
relativistically, so investigation of beamed ICCMB would be
inappropriate here.

Synchrotron particles with Lorentz factor $\gamma$ will upscatter CMB
photons ($\nu_{\rm CMB}$) to higher frequencies ($\nu_{\rm X}$) according to:
\begin{equation}
\frac{\nu_{\rm X}}{\nu_{\rm CMB}} = \displaystyle
\left(\frac{4}{3}\right)\gamma^2 - \frac{1}{3}.
\end{equation}
On the assumption that it is the photons from the peak of the CMB
distribution which are inverse Compton upscattered by the spent
synchrotron plasma in the lobes, then the Lorentz factors responsible
for the emission observed at 1\,keV are $1.5 \times 10^3$, so the
presence of ICCMB mandates the existence of $\gamma \sim 10^3$
particles in the lobe region.  Note that these particles will have
suffered considerable adiabatic expansion losses on leaving the
hotspot, but these preserve the shape of the spectrum.  The steep
power-law spectrum (if did continue to low energies) would predict
many more $\gamma \sim 10^3$ particles in the hotspot than in the
lobe, however, the absence of X-ray emission and thus ICCMB in the
east (left) hotspot mandates the absence of $\gamma \sim 10^3$
particles there, i.e.\ the particle spectrum in the hotspot cuts off
above these low energies.  If we conservatively take $\gamma_{\rm
min}$ in the lobe to be $\sim 10^3$ then this requires the lower limit
to the $\gamma_{\rm min}$ in the compact east hotspot to be $\sim
10^4$.

The X-rays are much more pronounced in the regions of the lobes
containing spent, rather than currently radiant, synchrotron plasma.
The extended X-ray emission is therefore most likely produced by
non-thermal particles which no longer contribute to the observed radio
emission.  In fact, the X-ray emission only seems to appear where the
east radio lobe has undergone significant transverse expansion (at
Right Ascension 09 08 21.5, see Figure 1); closer to the central
engine still, the lobe plasma has presumably undergone still further
transverse expansion so that its magnetic field strength is too low
there to cause the remaining low Lorentz factor particles to radiate
at radio wavelengths.  However, by Equation\,1, the X-ray emission
traces the location of old synchrotron particles, injected by the
hotspot a long time ago, whose energies are now reduced down to a
Lorentz factor $\gamma_{\rm min} \sim 10^3$.  The lobe radio emission
observed at 408\,MHz \citep{Law95} adjacent to the compact hotspots
will require rather higher Lorentz factors than this value of
$\gamma_{\rm min}$ (assuming magnetic field strengths $\ltsim\ 1\,{\rm
nT}$) as the particles recently accelerated in the hotspots and then
released into the lobes emit synchrotron radiation in the magnetic
fields there.  So whilst electrons with such low Lorentz factors in
the more aged regions of the lobes cannot produce synchrotron
radiation at observable radio frequencies, they are ideally matched to
upscatter CMB photons into the {\sl Chandra} X-ray energy bands.

The radio/X-ray comparison in Figure\,\ref{fig:overlay} shows a
separation of the outermost X-ray emission from the outermost radio
emission towards the east, by $\sim 80$\,kpc.  This is the first
direct indication of the elusive low-energy cutoff for plasma
energized in hotspots, as it implies that there are no such low-energy
particles in plasma recently accelerated in the hotspots.  This has
implications for the physical mechanism for particle acceleration in
hotspots: for example, it may be that the only particles to escape
from hotspots into lobes are those with gyro-radii larger than the
shock front, while less energetic particles with smaller gyro-radii
are very rapidly accelerated within the thickness of the shock (which
is of course on size-scales much smaller than current observations of
hotspots resolve).

The western hotspot of 6C\,0905+3955 appears to be an exception to
this behaviour, in that there is very approximate spatial
correspondence between the X-ray and radio emission (see inset to
Figure\,\ref{fig:overlay}).  It could be that in this side of the
source, X-ray emission is seen much closer to the radio hotspots
because expansion losses are so much greater in their vicinity (note
the transverse size of the west lobe is rather larger than that of the
east lobe) and lower-$\gamma$ particles are therefore expected to be
present much closer to this hotspot.  It is also possible that
different physics is responsible for this emission: for example,
synchrotron self-Compton as in the hotspots of Cygnus\,A
\citep{Wil01}, or perhaps X-ray emission from a galaxy responsible for
the putative gravitational lensing of this hotspot \citep{Law95}.

\section{Implications of a high $\gamma_{\rm min}$}
\label{sec:implications}
If the $\gamma_{\rm min}$ of the particles accelerated in the hotspots
is {\em higher} than has previously been thought, the equipartition
magnetic field strength, $B_{\rm eq}$ (calculated on the basis of the
observed luminosity and the assumption that in the lobes there is
equal energy stored in the magnetic fields as in the relativistic
particles), will be {\em lower} than previously thought.  The
relationship between $\gamma_{\rm min}$ and $B_{\rm eq}$ is given by:
\begin{equation}
  B_{\rm eq} \propto \displaystyle\left(\frac{1}{\gamma_{\rm
  min}}\right)^{\frac{2\alpha - 1}{3 + \alpha}}
\end{equation}
where $\alpha$ is the frequency spectral index of the flux density
$S_{\nu} \propto \nu^{-\alpha}$ at frequency $\nu$.  Thus, if
$\gamma_{\rm min}$ of the lobes is assumed to be 10 (or 100), but is
in reality $10^3$, then the inferred B-field is discrepantly larger
than the true value by 4 (or 2).  These are approximately the factors
by which estimated equipartition field strengths, calculated assuming
the lower values of $\gamma_{\rm min}$, are larger than independently
estimated B-fields from co-spatial inverse-Compton X-ray and
synchrotron radio emission \citep{Bru02,Har02,Kat03,Cro05}.  Thus,
knowledge that $\gamma_{\rm min}$ is higher than previously assumed
may well reconcile these discrepant values.

If lobe magnetic field strengths are {\em lower} than previously
assumed, a corollary is that a higher normalisation of the particle
energy spectrum is required to produce a given observed synchrotron
luminosity.  The number density normalisation $N_0$ of synchrotron
particles for a power-law distribution of particles depends on
$\gamma_{\rm min}$ as follows:
\begin{equation}
\label{eq:nnought}
  N_0 \propto \gamma_{\rm min}^\frac{(2\alpha-1)(1+\alpha)}{(3+\alpha)}, 
\end{equation}
where $\alpha$ is the frequency spectral index measured to be $\alpha
= 1.1$ in 6C\,0905+3955, giving $N_0 \propto \gamma_{\rm min}^{0.6}$ in
this object.  The total number density $n_{\rm rel}$ is given by
$n_{\rm rel} = N_0 / [2\alpha\,\gamma_{\rm min}^{2\alpha}]$.  

\section{Implications for the relativistic Sunyaev-Zeldovich effect}
\label{sec:sz}

The dependence of number density of particles on $\gamma_{\rm min}$
has implications for the relativistic Sunyaev-Zeldovich effect
\citep{Sun69,Zel70,Bir99} from the plasma lobes of radio galaxies.
The optical depth of relativistic electrons is given \citep{Ens00} by
\begin{equation}
\label{eq:opticaldepth}
  \tau_{\rm rel} = \sigma_{\rm T}\int{\rm d}\ell\ n_{\rm rel},
\end{equation}
where $\sigma_{\rm T}$ is the Thomson scattering cross-section and
$\ell$ is the line-of-sight distance through the lobe (taken to be
3\,arcsec which is 25\,kpc in the assumed cosmology) and $n_{\rm rel}$
is the number density of relativistic particles (obtained by
integrating over $N_0$ in Equation\,\ref{eq:nnought}, as stated at the
end of Section\,\ref{sec:implications}).  The fraction of photons
upscattered from the CMB via the relativistic Sunyaev-Zeldovich effect
is given by $\tau_{\rm rel}$ --- this upscattering is assumed to be at
the wavelength of the peak of the CMB photon distribution which at $z
= 1.88$ is 3.7\,mm, corresponding to a frequency of 82\,GHz.
6C\,0905+3955 is the target of future observations to verify the value
of $\gamma_{\rm min}$ found in this Letter to search for a decrement
of intensity in this frequency regime.

Taking a value of $10^3$ for the value of $\gamma_{\rm min}$ in the
lobes gives a value of $\tau_{\rm rel} \sim 10^{-10}$ for the radio
emitting regions.  The value of $\tau$ will obviously be increased
above this value in the regions where X-rays are observed.  If such
values of $\gamma_{\rm min}$ are typical (for $\alpha > 0.5$) then
radio galaxies are unlikely to be as important for CMB decrements as
suggested by \cite{Ens00}, unless the equipartition paradigm does not
in general hold in these objects.

Future experiments should be sensitive to such levels of contribution
to CMB decrements on the angular sizes probed by giant (or relic)
radio galaxies such as 6C\,0905+3955 and, in combination with future
Chandra imaging, deepen our understanding of the physical conditions
in these objects. 

\section{Summary}
\label{sec:summary}
The extended structures of the giant radio galaxy 6C\,0905+3955 in
X-rays and radio have been compared and show that the bulk of the
X-ray emission associated with non-compact components (i.e.\ lobe
material) preferentially occurs where the synchrotron plasma is
expected to be oldest.  Since X-ray emission from ICCMB comes from
electrons with Lorentz factor $\gamma \sim 10^3$, this strongly
suggests that there is an increased presence of $\gamma \sim 10^3$
particles in the older plasma than in the more recently accelerated
plasma.

If the equipartition paradigm is valid then, $\gamma_{\rm min} \gtsim\
10^3$ if generally true means that the equipartition magnetic field
strengths are lower than previously assumed.  The value of
$\gamma_{\rm min}$ affects the optical depth of the relativistic
electrons in the radio lobes; it is important to establish the
generality of our discovery in order to be sure about the importance
of the relativistic Sunyaev-Zeldovich effect from radio galaxy lobes
imprinted in the details of the CMB.

\vspace{1cm} 
KMB, CSC \& ACF thank the Royal Society for their support. We thank
Robert Laing, Peter Duffy \& Katrien Steenbrugge for helpful
discussions.  The VLA is a facility of the NRAO operated by AUI, under
co-operative agreement with the NSF. AC thanks the Italian MIUR and
INAF for support.

\begin{table*}
\centering
\begin{tabular}{cccc}
\hline
Region & Total counts & $\Gamma$  & $L_{\rm X}~\rm{(2:10\kev)}$ \\
       &(background)  &                   & ${10}^{44}$\ergps\ \\
\hline
nucleus      	    & $11$~~$(0.14)$  & $2$                   & $1.2$ \\
lobe         	    & $81$~~$(27.19)$ & ${2.7}_{-0.7}^{+0.7}$ & $2.1$ \\
                    &                 & $2$                   & $2.8$ \\
western-most X-rays & $15$~~$(2.4)$   & ${2.5}_{-0.6}^{+0.7}$ & $0.6$ \\
                    &                 & $2$                   & $0.7$ \\
eastern-most X-rays & $9$~~$(1.57)$   & ${1.4}_{-0.3}^{+1.0}$ & $0.5$ \\
                    &                 & $2$                   & $0.4$ \\
\hline
\end{tabular}
\caption{The photon numbers for detection purposes are from the 0.3 to
  3\kev\ band while the spectral fits were made over the full 0.5 to
  7\kev\ band.  For the eastern-most X-ray emitting region, which is
  some 80\,kpc\ from the east radio hotspot, the probability of
  detecting 9 photons when 1.57 is expected is $3.9\times 10^{-5}$, so
  its detection is very secure. C-statistics were used to calculate
  the X-ray photon index, $\Gamma$, and the X-ray luminosity in the
  rest-frame $2-10$\kev -band, $L_{\rm X}$. The first, third, fifth
  and seventh lines in the table list the luminosity calculated for a
  fixed photon index of 2.}
\label{tab:results}
\end{table*}

\onecolumn


\begin{thebibliography}{10}

\bibitem[Baldwin(1982)]{Bal82} Baldwin, J.~E.\ 1982, IAU 
Symp.~ 97: Extragalactic Radio Sources, 97, 21 

\bibitem[Birkinshaw(1999)]{Bir99}
{Birkinshaw}, M., (1999) Phys. Rep., 310, 97

\bibitem[Blundell, Rawlings \& Willott(1999)]{Blu99}
{Blundell}, K.~M., {Rawlings}, S.  \& {Willott}, C.~J, 1999, \aj, 
117, 677

\bibitem[Blundell \& Rawlings(2000)]{Blu00}
{Blundell}, K.~M.\ \& {Rawlings}, S., 2000, \aj, 
119, 1111

\bibitem[Brunetti et al(2002)]{Bru02}
{Brunetti}, G., {Bondi}, M., {Comastri}, A.  \& {Setti}, G., 2002, \aap, 381, 795

\bibitem[Carilli et al(1991)]{Car91}
{Carilli}, C.~L., {Perley}, R.~A., {Dreher}, J.~W.  \& {Leahy}, J.~P.,
1991, \aj, 383, 554

\bibitem[Celotti et al(1998)]{Cel98}
{Celotti}, A., {Kuncic}, Z., {Rees}, M.~J.  \& {Wardle}, J.~F.~C.,
1998, MNRAS, 293, 288

\bibitem[Celotti(2003)]{Cel03}
{Celotti}, A., 2003,  Ap\&SS, 288, 175

\bibitem[Condon et al(1998)]{Con98}
Condon, J.~J., Cotton, 
W.~D., Greisen, E.~W., Yin, Q.~F., Perley, R.~A., Taylor, G.~B., \& 
Broderick, J.~J.\ 1998, \aj, 115, 1693 

\bibitem[Croston et al(2005)]{Cro05}
{Croston}, J.~H., {Hardcastle}, M.~J., {Harris}, D.~E., {Belsole}, E.,
  {Birkinshaw}, M.\  \& {Worrall}, D.~M., 2005, ApJ, 626, 733

\bibitem[Ensslin \& Kaiser(2000)]{Ens00}
Ensslin, T.~A., \& Kaiser, C.~R.\ 2000, \aap, 360, 417 

\bibitem[Fanaroff \& Riley(1974)]{Fan74} Fanaroff, B.~L., \& 
Riley, J.~M.\ 1974, \mnras, 167, 31P 

\bibitem[Ficarra, Grueff \& Tomassetti(1985)]{Fic85}
Ficarra, A., Grueff G.\ \& Tomassetti G., 1985, A\&AS, 59, 255

\bibitem[Hardcastle(2005)]{Har05} 
Hardcastle, M.~J.\ 2005, \aap, 434, 35 

\bibitem[Hardcastle, Birkinshaw \& Worrall(2001)]{Har01}
{Hardcastle}, M.~J., {Birkinshaw}, M.  \& {Worrall}, D.~M., 2001, MNRAS, 323, L17

\bibitem[Harris \& Krawczynski(2002)]{Har02}
{Harris}, D.~E. \& {Krawczynski}, H., 2002,  \apj, 565, 244

\bibitem[Kaiser et al.(1997)]{Kai97} 
Kaiser, C.~R., Dennett-Thorpe, J., \& Alexander, P.\ 1997, \mnras, 292, 723 

\bibitem[Kataoka et al(2003)]{Kat03}
{Kataoka}, J., {Leahy}, J.~P., {Edwards}, P.~G., {Kino}, M., {Takahara}, F.,
  {Serino}, Y., {Kawai}, N.  \& {Martel}, A.~R., 2003, \aap, 410, 833

\bibitem[Law-Green et al(1995)]{Law95}
{Law-Green}, J.~D.~B., {Eales}, S.~A., {Leahy}, J.~P., {Rawlings}, S.  \&
  {Lacy}, M., 1995, MNRAS, 277, 995

\bibitem[Miley(1980)]{Mil80} Miley, G.\ 1980, \araa, 18, 165 
 
\bibitem[Scheuer \& Williams(1968)]{Sch68}
{Scheuer}, P.~A.~G. \& {Williams}, P.~J.~S., 1968, ARAA,  6, 321

\bibitem[Sunyaev \& Zeldovich(1969)]{Sun69}
{Sunyaev}, R.~A. \& {Zeldovich}, Y.~B., 1969, Nature, 223, 721

\bibitem[Wardle et al(1998)]{War98}
{Wardle}, J.~F.~C., {Homan}, D.~C., {Ojha}, R.  \& {Roberts}, D.~H.,
1998, Nature, 395, 457

\bibitem[Wilson, Young \& Shopbell(2001)]{Wil01}
{Wilson}, A.~S., {Young}, A.~J.  \& {Shopbell}, P.~L., 2001, \apj, 547, 740

\bibitem[Zeldovich \& Sunyaev(1970)]{Zel70}
{Zeldovich}, Y.~B. \& {Sunyaev}, R.~A., 1970, ApSS, 7, 20

\end{thebibliography}
\end{document}